\documentclass{naturefig}
\usepackage{graphicx}
\usepackage{amssymb}
\RequirePackage{cite}
\RequirePackage{times}
\RequirePackage{fullpage}
\RequirePackage{ifthen}

\title{\textbf{Gas accretion as the origin of chemical abundance gradients in distant galaxies }}

\author{G. Cresci$^{1,2}$, F. Mannucci$^1$, R. Maiolino$^3$, A. Marconi$^4$, A. Gnerucci$^4$ \& L. Magrini$^1$}

\begin{document}

\maketitle

\begin{affiliations}
 \item INAF - Osservatorio Astrofisico di Arcetri, Largo E. Fermi 5, 50125, Firenze, Italy
 \item Max-Planck-Institut f\"ur extraterrestrische Physik (MPE), Giessenbachstr. 1, 85748, Garching, Germany
 \item INAF - Osservatorio Astronomico di Roma, via di Frascati 33, 00040, Monte Porzio Catone, Italy
 \item Dipartimento di Fisica e Astronomia, Universit\'a di Firenze, Largo E. Fermi 2, 50125, Firenze, Italy
\end{affiliations}

\begin{abstract}
It has recently been suggested \cite{dekel09} \cite{borneaud09} that galaxies in the early Universe can grow through the accretion of cold gas, and that this may have been the main driver of star formation and stellar mass growth \cite{fs09} \cite{daddi07} \cite{tacconi10}. Because the cold gas is essentially primordial, it has a very low abundance of elements heavier than helium (metallicity). As it is funneled to the centre of a galaxy, it will lead the central gas having an overall lower metallicity than gas further from the centre, because the gas further out has been enriched by supernovae and stellar winds, and not diluted by the primordial gas. Here we report chemical abundances across three rotationally-supported star-forming galaxies at z$\sim$3, only 2 Gyr after the Big Bang. We find an 'inverse' gradient, with the central, star forming regions having a lower metallicity than less active ones, opposite to what is seen in local galaxies \cite{garnett97} \cite{magrini09}. We conclude that the central gas has been diluted by the accretion of primordial gas, as predicted by 'cold flow' models.
\end{abstract}

\begin{figure}
\centerline{\includegraphics[width=0.5\textwidth]{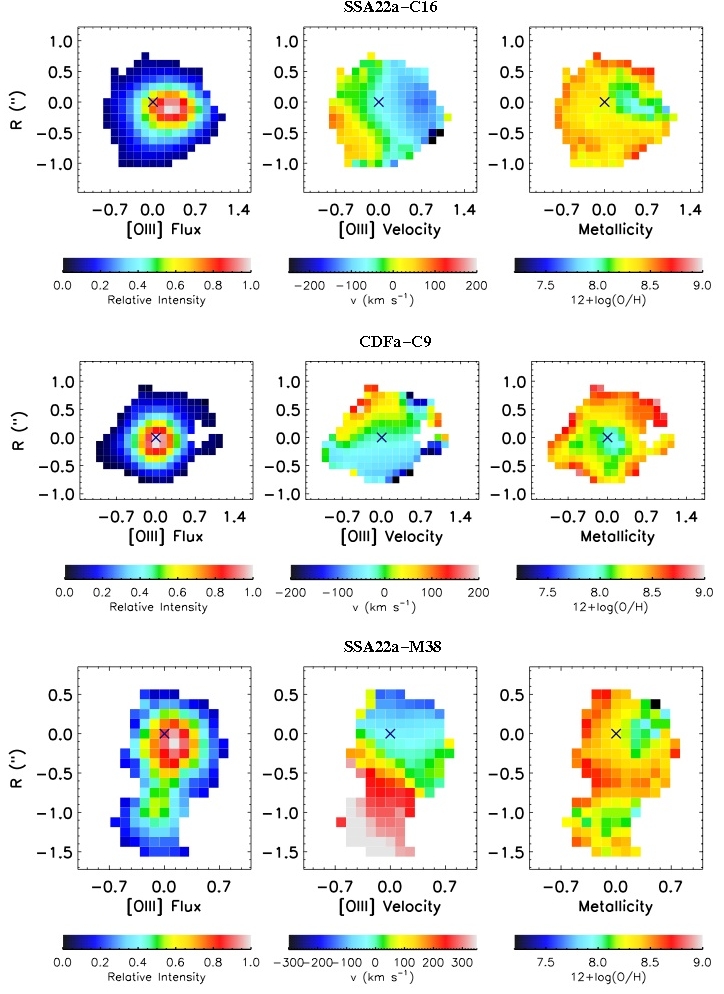}}
\caption{\footnotesize{Surface brightness and velocity of the [OIII]$\lambda$5007 line, and metallicity maps. The data for SSA22a-C16 ($z=3.065$, \textit{upper panels}), CDFa-C9 ($z=3.219$, \textit{central panels}) and SSA22a-M38 ($z=3.288$, \textit{lower panels}) were obtained with the SINFONI spectrograph using the $0.125\times0.250''$ pixel scale in seeing-limited conditions, resulting in a spatial resolution of $\sim 0.5''$ (Full Width Half Maximum of the Point Spread Function, PSF). The maps were extracted from the SINFONI datacube after a Gaussian smoothing with FWHM=3 pixels ($0.375''$). The \textit{left panels} in each row show normalized surface brightness in the [OIII]$\lambda$5007 emission line. The same line has been used to derive the velocity maps shown in the \textit{central panels}; the observed gas kinematics is compatible with a rotating disk, with no evidence of merger-induced complex dynamics. The \textit{right panels} show maps of gas phase metallicity, as relative abundances of oxygen and hydrogen parameterized in units of $12+log(O/H)$. Lower metallicity region, corresponding to a higher [OIII]/[OII] ratio, are surrounded by a more enriched disk. The crosses in each panel mark the position of the continuum peak.}}
\label{figmet}
\end{figure}

We selected three Lyman-break galaxies \cite{steidel03} among the AMAZE \cite{maiolino08} and LSD \cite{mannucci09} sample, SSA22a-C16 ($z=3.065$), CDFa-C9 ($z=3.219$) and SSA22a-M38 ($z=3.288$), to test the presence of metallicity gradients in galaxies beyond $z=3$, when the Universe was only 14-16\% of its actual age, and only a small fraction ($\sim5-10\%$ \cite{pozzetti07}) of the total stellar mass in the Universe had been assembled. This epoch is particularly interesting, as it precedes the peak of the cosmic star formation density \cite{hopkins06}. We selected these three galaxies because they show a remarkably symmetric velocity field in the [OIII] emission line, which traces the ionized gas kinematics (see Fig. \ref{figmet}). Such kinematics indicates that these are rotationally supported disks (Gnerucci et al., in preparation), with no evidence for more complex merger-induced dynamics. Assuming an exponential disk mass model to reproduce the dynamical properties of the three sources, we derive dynamical masses in the range $\sim10^{10}-2\ 10^{11}\ M_{\odot}$. By using the H$\beta$ line flux \cite{kennicutt98a}, we infer star formation rates as high as $\sim 120\ M_{\odot}/yr$ after correcting for dust extinction \cite{maiolino08}. 
We carefully checked for the presence of Active Galactic Nuclei (AGN) in our galaxies, as they would invalidate the calibrations derived for star-forming galaxies because of the different ionizing continua involved (see Supplementary information). 

Near-infrared spectroscopic observations of the galaxies were obtained with the integral field spectrometer SINFONI \cite{eisenhauer03} on the Very Large Telescope of the European Southern Observatory. 
The spectra cover the $1.45-2.41 \mu m$ wavelength range; strong rest frame optical emission lines of the observed galaxies, such as [OIII]$\lambda$5007, [OIII]$\lambda$4959, H$\beta$, [NeIII]$\lambda$3870 and [OII]$\lambda$3727, are redshifted to this range.
The flux ratio between the (rest-frame) optical nebular lines observed in SINFONI spectra depend on gas metallicity, 
and have been calibrated for local low and high metallicity galaxies \cite{nagao06}. In particular, we use a combination of three independent metallicity diagnostics \cite{maiolino08}, [OIII]$\lambda$5007/H$\beta$, [OIII]$\lambda$5007/[OII]$\lambda$3727 and, when available, [NeIII]$\lambda$3870/[OII]$\lambda$3727, to derive the gas phase metallicity in our spectra. Even if a single line ratio can constraint the metallicity in a broad interval of $\sim 0.4$ dex or more, the use of three independent diagnostics helps us in removing the degeneracies and reducing the uncertainties; we can  determine for each spatial pixel the combination of metallicity and extinction that best fit the observed line ratios with our calibrations (see Supplementary information for the details).

The derived metallicity maps are shown in Fig. \ref{figmet}. An unresolved region with lower metallicity is evident in each map, surrounded by a more uniform disk of higher metal content. In CDFa-C9 the lower metallicity region is coincident with the galaxy center, as traced by the continuum peak, while it is offset by $\sim 0.60''$ (4.6 kpc) in SS22a-C16 and $\sim0.45''$ (3.4 kpc) in SS22a-M38. On the other hand, in all the galaxies the area of lower metallicity is coincident or closer than $0.25''$ (1.9 kpc, which is half the point spread function full-width at half-maximum) to the regions of enhanced line emission, which trace the more active star forming regions.
The average difference between high and low metallicity regions in the three galaxies is  $0.55$ in units of 12+log(O/H), larger than the $\sim0.2-0.4$ dex gradients measured in the Milky Way and other local spirals \cite{vanzee98} on the same spatial scales. The measured gas phase abundance variations have a significance between 98\% and 99.8\% (see Tab.~\ref{properties}).

The observed line ratios depend on metallicity both directly and indirectly through the ionization parameter $U$
(the ratio between the flux of ionizing photons and the gas density), with lower metallicity star forming regions generally have higher U \cite{nagao06}.  In principle, the observed variation of line ratios across galaxies could still be due to
variations of $U$ unrelated to the metallicity distribution. As local galaxies show a broad range of ionization parameter, this dependence on U is responsible for the spread included in our calibrations. In order to exclude the possibility that the line ratios in our galaxies are biased due to significantly different values of ionization parameter at high redshift, we have examined their dependence on $U$ based on the latest photoionization models available \cite{dopita06} \cite{levesque10} \cite{martinmanjon10}.
These models are not able to simultaneously reproduce the different observed line ratios with a variation of the ionization parameter alone: at constant metallicity the observed line ratios differ from models predictions by $\sim0.2$ dex. 
This demonstrate that different metallicities are required to reproduce the observations, and even if other effects are present, they are not dominant. 

Observed metallicity gradients in local disk galaxies show metallicities decreasing outward from the galactic centers. In contrast, our high redshift sources show ``inverted'' gradients, with lower metallicity in a well defined active region closer to the galaxy center. Current models of chemical enrichment in galaxies \cite{molla97} \cite{hou00} cannot reproduce our observations at the moment, as they assume radially isotropic gas accretion onto the disk and the instantaneous recycling approximation \cite{tinsley79}. Nevertheless, the detected gradients can be explained in the framework of the cold gas accretion scenario  \cite{keres05}  recently proposed to explain the properties of gas rich, rotationally supported galaxies observed at high redshift \cite{cresci09} \cite{fs09}. In this scenario, the observed low metallicity regions are created by the local accretion of metal-poor gas in clumpy streams \cite{dekel09}; this gas penetrate deep into the galaxy following the potential well, and sustaining the observed high star formation rate in the pre-enriched disk. Stream-driven turbulence is then responsible for the fragmentation of the disks into giant clumps, as observed at $z \geq 2$ \cite{genzel08} \cite{mannucci09}, that are the sites of efficient star formation and possibly the progenitors of the central spheroid. This scenario is also in agreement with the dynamical properties of our galaxies, which appear to be dominated by gas rotation in a disk with no evidence of the kinematical asymmetries typically induced by mergers .

\begin{figure}
\centerline{\includegraphics[width=0.5\textwidth]{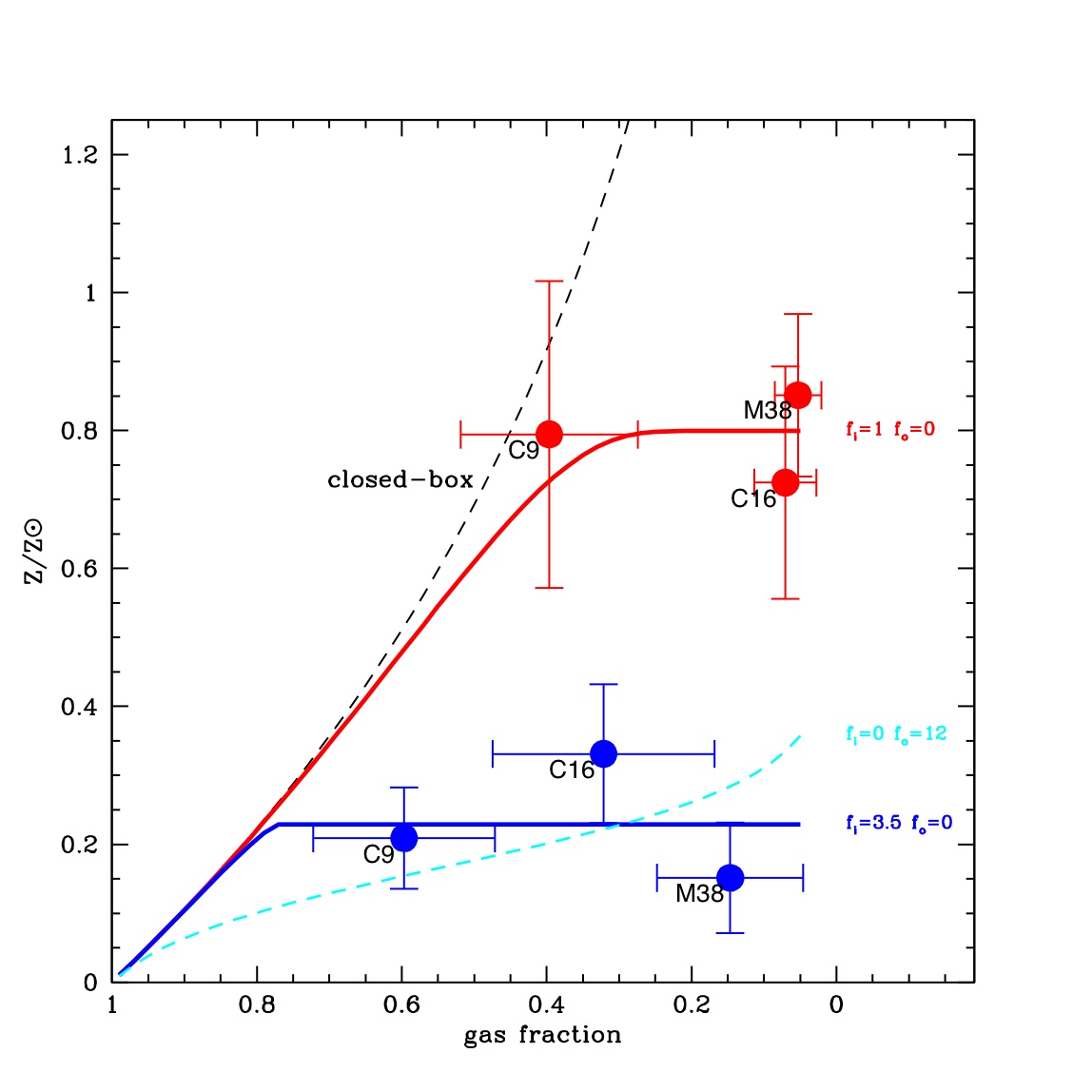}}
\caption{\footnotesize{Gas fractions for the different regions in the observed galaxies. To estimate gas masses in the low metallicity regions and in the rest of the galaxies, we used the H$\beta$ flux corrected for extinction. We convert measured star formation rate densities from H$\beta$ to a gas mass by inverting the Kennicutt-Schmidt law \cite{kennicutt98b}. To estimate stellar masses, we have to relay on the continuum flux from our SINFONI data, as we are still missing deep and high resolution continuum images of the sources. In the datacubes of our galaxies the continuum is detected, although at a lower S/N than the emission lines, making this estimate more uncertain. We have to further assume that the stellar mass scales as the K-band continuum light (V band rest frame). We thus measure the total K band continuum flux on the whole galaxy and the flux in the lower metallicity regions, using these relative flux measurements to rescale the total stellar mass from Spectral Energy Distribution fitting \cite{maiolino08}, and estimate the stellar mass in the metal poor and metal rich regions. We plot the derived gas fraction against the metallicity in solar units $Z/Z_{\odot}$. Filled blue and red circles show respectively the low and high metallicity regions of each galaxy; black dashed line, expectations from a closed-box model. The difference between the regions can be explained by a simple model including only infall of gas $f_i$, which is, respectively, 3.5 (blue solid line) and 1 times (red solid line) the SFR for the regions of lower and higher metallicity. Dashed cyan lines, a model of pure outflow $f_o$, 12 times the SFR. The observed difference in gas fraction it is consistent with the expectation from a massive infall of low metallicity gas in the central regions. The error bars are $1 \sigma$ formal uncertainties on the measuraments, given the assumptions discussed.}}
\label{figyields}
\end{figure}

If infall and outflows are not present, metallicity is directly related to the fraction of gas present in the galaxy. For this reason additional hints on the presence of infall and outflows can be obtained considering these two quantities.  
We can derive an estimate of gas fractions in each single region of our galaxies assuming that the relation between SFR and gas density observed in the local universe (the Schmidt-Kennicutt relation \cite{kennicutt98b}) also 
applies to high redshift galaxies \cite{bouche07}. The gas fractions obtained for the two regions in each galaxy are shown in Fig.~\ref{figyields} plotted againts metallicities $Z$. It is evident that our results do not support the closed box model (black line)
where galaxies have no contribution from infalls and outflows.
Our data can be reproduced only by including either outflows of very enriched material, such as the ejecta from SNe, which extract
metals and decrease $Z$, or infalls of pristine gas which increase the gas fraction and decrease $Z$.
 In particular, if strong infall of pristine gas is present, we would expect higher gas fractions in the low metallicity regions, as observed. Using simple models\cite{erb08}\cite{mannucci09}, we estimate the amount of infall and outflow required to reproduce the observations. We take into account two opposite scenarios. If inflows are dominant, we require $\sim3.5$ times more pristine gas infall in the low metallicity regions than in the high metallicity part. On the other hand, if outflows are dominant, high metallicity regions can be explained with outflow $\sim2$ times the SFR, while unrealistic outflows $\sim12$ times the SFR would be needed to explain the properties of low metallicity regions. We therefore conclude that the differences in gas fractions as derived by our estimates are not only reasonable given the measured metallicity differences, but are also consistent with expectations from massive infalls of low metallicity gas in the central regions of the galaxies. They are unlikely to be explained using outflows only.

Our observations of low metallicity regions in three galaxies at $z\sim3$ therefore provide the evidence for the actual presence of accretion of metal-poor gas in massive high-z galaxies, capable to sustain high star formation rates without frequent mergers of already evolved and enriched sub-units. This picture was already indirectly suggested by recent observational studies of gas rich disks at $z\sim1-2$\cite{fs09}\cite{tacconi10}, by the study of the variation of metallicity as a function of gas fraction \cite{mannucci09} and star formation rate \cite{mannucci10}. The data presented here are now directly revealing the metal poor accreted gas responsible for fueling the star formation rate in massive galaxies in the early Universe.

Although the three galaxies discussed here are the only $z\sim3$ sources for which a metallicity gradient has been measured at the moment, the detection of lower metallicity clouds in all three galaxies suggests that metal poor gas accretion is probably a common feature in actively rotationally supported star forming galaxies at this early cosmic epoch. However, it is hard to speculate how much this accretion mechanism is dominant in the global star forming population, as the relative number of rotationally supported and more dynamically complex galaxies is still unknown, and similar observations are still missing in different classes of galaxies.


\begin{addendum}
 \item SINFONI data were obtained with observations made with 
the ESO Telescopes at the Paranal Observatories. 
We would like to thank the ESO staff for their work and support. This work has been supported by INAF and ASI. 
 \item[Authours contributions] All authors have contributed extensively to data reduction and interpretation.
 \item[Competing Interests] The authors declare that they have no competing financial interests.
 \item[Correspondence] Correspondence and requests for materials should be addressed to G.C.~(email: 
gcresci@arcetri.astro.it).

\end{addendum}

\newpage

\begin{table*}[h]
	\begin{center}
	\begin{small}
	\begin{tabular}{l c c c c c}
	\hline
	\hline
	Name & $z$ & log($M_*/M_{\odot}$) & 12+log(O/H) & 12+log(O/H) & 12+log(O/H) \\
	     & & & (tot) & (low Z) & (high Z) \\
	\hline
	SSA22a-C16 & 3.065 & $10.68^{+0.16}_{-0.54}$ & $8.36^{+0.06}_{-0.06}$ & $8.18^{+0.13}_{-0.14}$ & $8.52^{+0.14}_{-0.07}$ \\
	CDFa-C9    & 3.219 & $10.03^{+0.40}_{-0.08}$ & $8.36^{+0.06}_{-0.05}$ & $7.98^{+0.14}_{-0.16}$ & $8.56^{+0.12}_{-0.12}$ \\
	SSA22a-M38 & 3.288 & $10.86^{+0.18}_{-0.41}$ & $8.26^{+0.09}_{-0.11}$ & $7.84^{+0.22}_{-0.23}$ & $8.59^{+0.05}_{-0.07}$ \\
	\hline
	\hline
	\end{tabular}
	\end{small}
	\caption{Properties of the sample galaxies. $z$ is the source redshift, $M_*$ the stellar mass derived from multi band photometry and spectral energy distribution fitting assuming a Chabrier \cite{chabrier03} Initial Mass Function \cite{maiolino08}, while the last three columns show the metallicity integrated for the whole galaxy, for the low metallicity regions and for the high metallicity regions respectively, in units of 12+log(O/H). The amplitude and significance of the detected gradients were evaluated extracting 1D spectra integrated over apertures comparable to the PSF (FWHM=$0.5''$) in both the lowest and highest metallicity region, and using them to measure the line ratios (see Supplementary information). The uncertainties are the 1$\sigma$ confidence intervals. }
	\label{properties}
	\end{center}
\end{table*}

\newpage

\section*{Supplementary Information} 

\noindent
\textbf{Measuring metallicities:} At the redshift of the galaxies in our sample ($z=3.0-3.3$), five prominent rest-frame optical emission lines are observable in the H+K SINFONI grism: [OII]$\lambda$3727 and [NeIII]$\lambda$3870 in the H band, H$\beta$, [OIII]$\lambda$4959 and [OIII]$\lambda$5007 in the K band. The resulting line ratios allow us to measure the gas phase metallicity using five well calibrated \textsuperscript{15} 
diagnostics, $R_{23}$ = ([OII]$\lambda$3727 + [OIII]$\lambda$4959 + [OIII]$\lambda$5007)/H$\beta$ \cite{2pagel79}, [OIII]$\lambda$5007/H$\beta$, [OII]$\lambda$3727/H$\beta$, [OIII]$\lambda$5007/[OII]$\lambda$3727 and [NeIII]$\lambda$3870/[OII]$\lambda$3727. 
Although these calibrations were obtained in local galaxies, where the physical conditions, such as the ionization parameter, might be different respect to higher redshift \cite{2brinchmann08}, these strong line metallicity diagnostics do not deviate at high-z more than about 0.1 dex \cite{2liu08}. 

These calibrations only apply if the ionizing source is stellar radiation, while the presence of an AGN ionizing continuum would invalidate the calibrations derived for star-forming galaxies. This could be a problem for our sources, as the larger metallicity differences are detected close to the galaxy nucleus, where an eventual AGN contribution would be more important.
We therefore checked that, for our galaxies, rest frame UV spectra, soft and hard X-ray data, as well as $24 \mu m$ Spitzer data show no indication of AGN \textsuperscript{9}. 
Moreover, we verified with spectral classification diagrams \cite{2lamareille10} that the line ratios detected in the central nuclear regions of our sources are compatible with a pure starburst in a low metallicity galaxy, ruling out the presence of any AGN contribution to the detected gradients, or significantly different physical conditions in the star forming regions with respect to local galaxies. 

\begin{figure}
\centerline{\includegraphics[angle=0,width=0.8\textwidth]{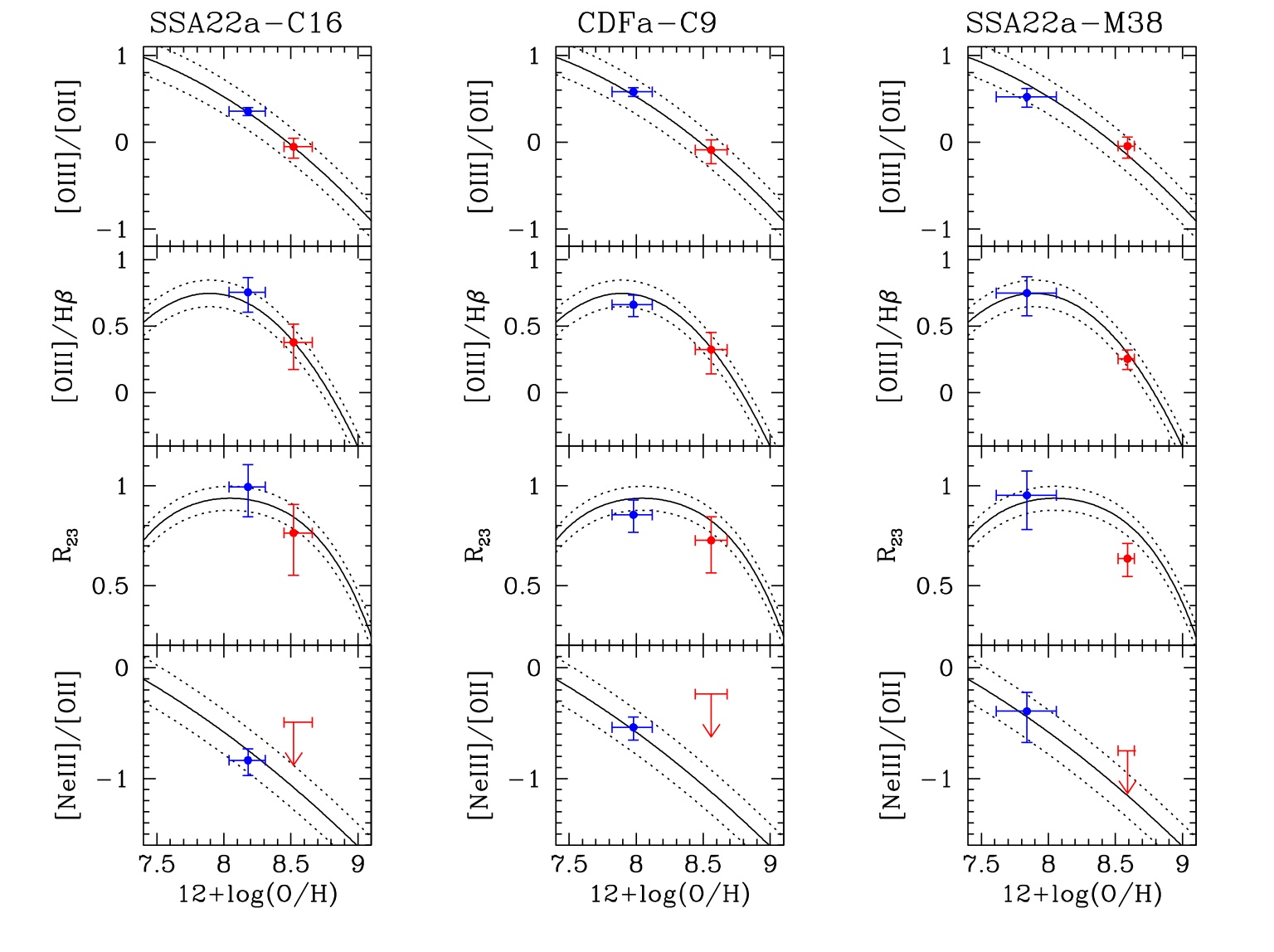}}
\caption{Diagnostic tools used to measure metallicities. The low and high metallicity regions for the three galaxies are plotted, SSA22a-C16 on the left, CDFa-C9 in the center, SSA22a-M38 on the right. The \textit{first panels} show the [OIII]$\lambda$5007/[OII]$\lambda$3727 line ratio as a function of metallicity, the \textit{second panels} show the [OIII]$\lambda$5007/H$\beta$, that is very similar to $R_{23}$ (shown in the \textit{third panels} but not used for the minimization) at low metallicities, where the [OII]$\lambda$3727 flux is small compared to [OIII]. The \textit{fourth panels} show the [NeIII]$\lambda$3870/[OII]$\lambda$3727 ratio. The blue points are relative to the lower metallicity regions, and the red points to the higher metallicity ones. The errorbars denotes the 1~$\sigma$ uncertainties in the flux measurements or in metallicity fitting. An upper limit is shown where the [NeIII] emission line was not detected.}
\label{exfit}
\end{figure}

\begin{figure}
\centerline{\includegraphics[width=0.65\textwidth]{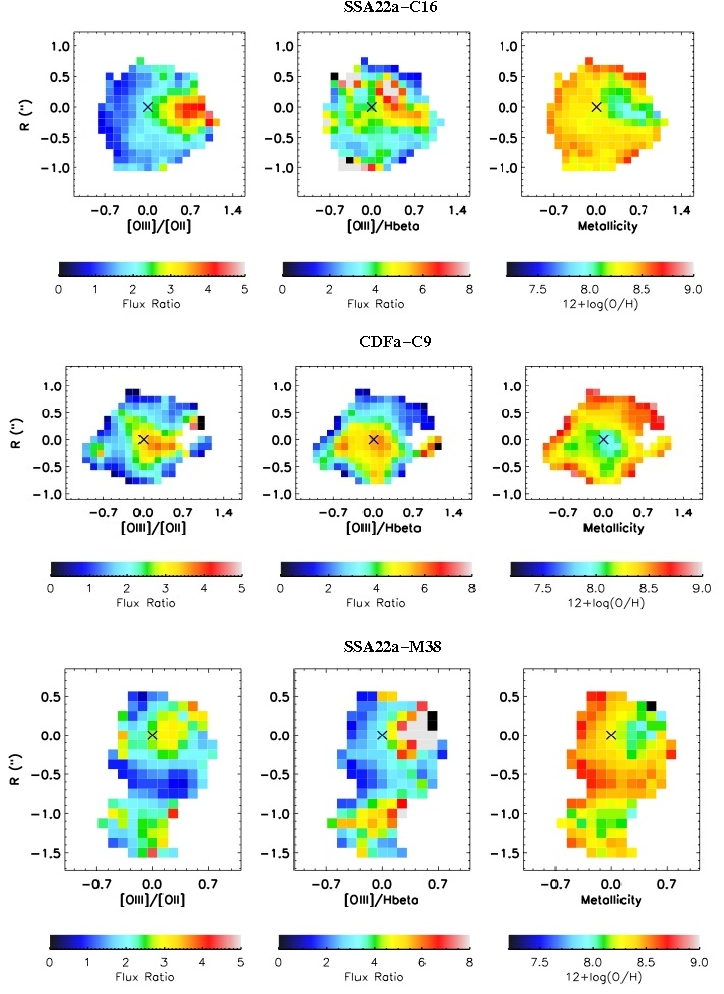}}
\caption{Line ratios and metallicity maps for the three galaxies of the sample. These maps are obtained after a Gaussian smoothing with FWHM=3 pixels ($0.375''$). The \textit{left panels} show the [OIII]$\lambda$5007/[OII]$\lambda$3727 line ratio maps, while the [OIII]$\lambda$5007/H$\beta$ maps are presented in the \textit{central panels}. The line ratios gradients have a significance between $3$ and $5\sigma$. The \textit{right panels} show the resulting maps of gas phase metallicities, as relative abundances of oxygen and hydrogen parameterized in units of $12+log(O/H)$. The crosses in each panel mark the position of the continuum peak.}
\label{figrat}
\end{figure}

Obviously, not all of the five diagnostics listed above are independent, and each has its own advantages and disadvantages. [OIII]$\lambda$5007/H$\beta$ is essentially unaffected by dust extinction, but it allows two different metallicities for a given ratio (see Fig.~\ref{exfit}). On the other hand, [OIII]$\lambda$5007/[OII]$\lambda$3727 has a monotonic dependence on metallicity, but it is affected by reddening. [NeIII]$\lambda$3870/[OII]$\lambda$3727 is instead both monotonic and less affected by dust, but the [NeIII]$\lambda$3870 line is much fainter than the others, and thus not always observable. However, the combination of these three indicators allows to measure unambiguously the metallicity, taking into account the effect of dust reddening (see Fig.~\ref{figrat}). We exclude the remaining two line ratios from the following analysis, as $R_{23}$ is very similar to [OIII]$\lambda$5007/H$\beta$ at low metallicities and [OII]$\lambda$3727/H$\beta$ is not independent from the others. 

\begin{figure}
\centerline{\includegraphics[width=0.65\textwidth]{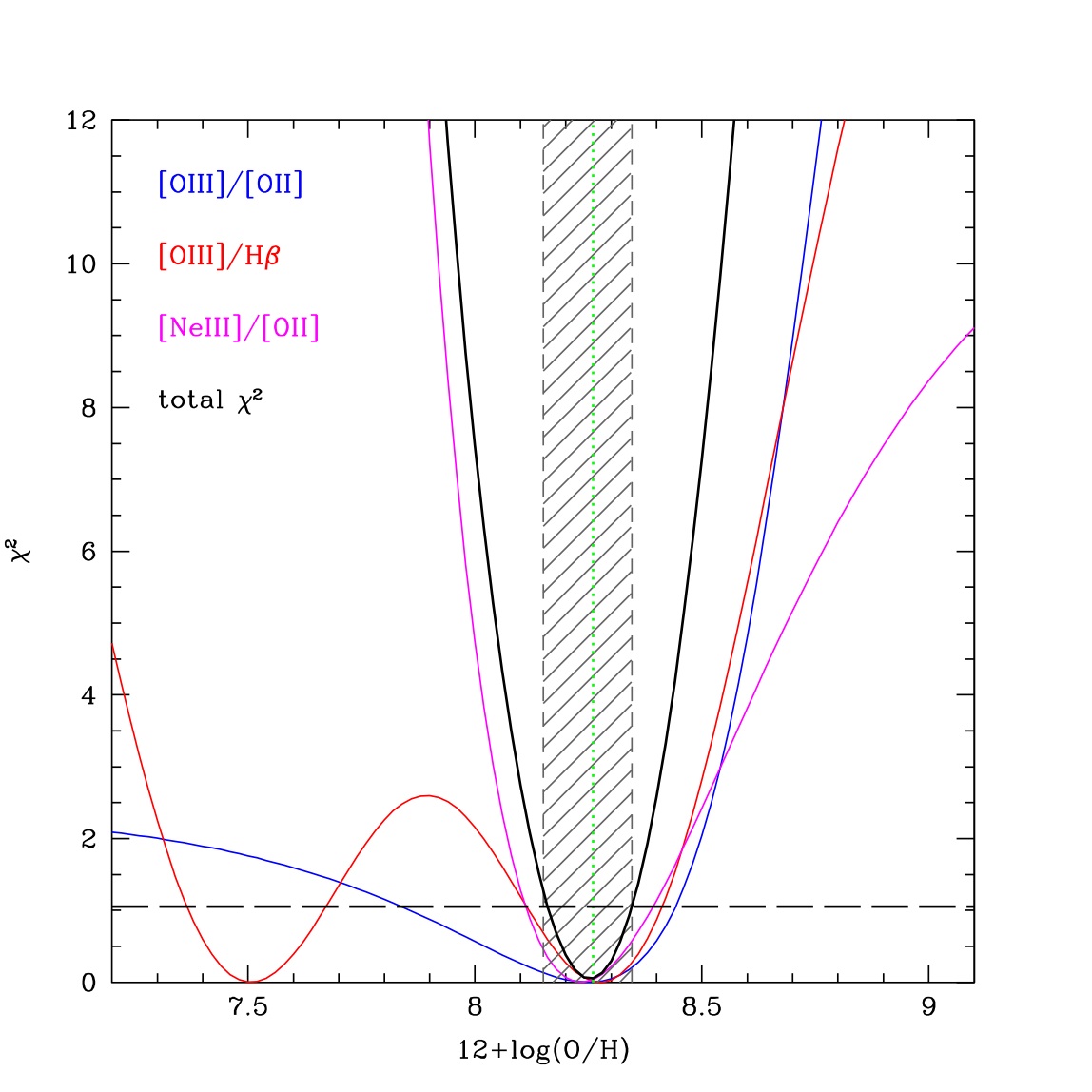}}
\caption{Example of $\chi^2$ dependence on metallicity using three independent diagnostics. The variation of $\chi^2$ as a function of metallicity is shown as an example for the integrated spectra of SSA22a-M38 using the [OIII]/[OII] line ratio only (blue curve), [OII]/H$\beta$ only (red curve), and [NeIII/[OII] only (magenta curve). The total $\chi^2$ from the combined three independent measurements is shown as a black curve: the global 1 $\sigma$ confidence interval, represented as a gray shaded region is significantly narrower in this case. The best fitting value is marked with a green dotted line.}
\label{chi2}
\end{figure}

For each spatial pixel of the datacube with $S/N\geq3$ in [OIII]$\lambda$5007/H$\beta$, we determine the best combination of metallicity and extinction that minimizes the $\chi^2$ in the three selected diagnostics, after a Gaussian smoothing of the data with a FWHM=3 pixels ($0.375''$, lower than the spatial resolution of $\sim0.5''$). To compute the $\chi^2$, we take into account both the uncertainties on the line ratio measurements and the dispersion of each calibration \textsuperscript{15} 
Such a dispersion is mostly due to the broad range of ionization parameters in the galaxies used to derive the calibrations\textsuperscript{15}. 
There is no indication that the metallicity calibrations and the spread based on local galaxies may fail at high redshift, because the measured line ratios for our galaxies are consistent with the ones measured in local galaxies, and the comparison with the photoionization models available do not support the presence of unusually high ionization parameters.
The spread in the calibrations results in a large uncertainty when the metallicity is measured using a single line ratio ($\sim0.2$ dex or more, depending on the line ratio), but the combination of the three independent tracers helps us to decrease the formal uncertainty (see Fig.~\ref{chi2}) . The additional systematic uncertainty on the {\it absolute} metallicity associated with the photoionization models used to derived the calibrations is not considered, as we are only interested in relative metallicity differences. The fitting results are shown in Fig.~\ref{figrat} for the three galaxies. 

To quantify the significance of the detected gradients we extracted 1D spectra integrated over circular apertures within a diameter of $0.5''$, comparable to the PSF. We selected two apertures in each galaxy, one centered on the lowest metallicity region and one with highest metallicity, and we used the resulting line ratios to derive the metallicities in both regions. The fits are shown in Fig.~\ref{exfit} for the three galaxies, and the results are listed in Tab.~1. The allowed values for the dust extinction are in the range $0<A_V<1.2$, in agreement with the estimate derived from the SED fitting $A_V\sim0.7-1$ for the three galaxies\textsuperscript{9}. 
Although the extinction has a large uncertainty, the metallicity is relatively well constrained.

\noindent
\textbf{Supplementary References}


\end{document}